\documentclass[a4paper]{iopart}
\usepackage{epsfig}
\usepackage{iopams}

\newcommand{\be} {\begin {equation}}
\newcommand {\ee} {\end {equation}}
\newcommand{\ba} {\begin {array}}
\newcommand {\ea} {\end {array}}
\newcommand{\bea}{\begin{eqnarray}}
\newcommand{\eea}{\end{eqnarray}}

\newcommand{\g}{\mathfrak{g}}

\newcommand{\Q}{\mathcal{Q}}

\newcommand{\C} {\mathbb{C}^N}
\newcommand{\CP} {\mathbb{C}P^{N-1}}

\newcommand{\p}{\partial_+}
\newcommand{\bp}{\partial_-}

\newcommand{\0}{{\bf \varnothing} }
\newcommand{\Pb}{{ P}}

\renewcommand{\H}{{ H} }
\newcommand{\F}{{ F} }

\newtheorem{theorem}{Theorem}

\begin{document}

\title{Soliton surfaces associated with $\CP$ sigma models}
\author{ A M Grundland$^1$ $^2$ and S Post$^1$}

\address{$^1$ Centre de Recherches Math\'ematiques. Universit\'e de Montr\'eal. Montr\'eal CP6128 (QC) H3C 3J7, Canada}
\address{$^2$ Department of Mathematics and Computer Sciences, Universit\'e du Quebec, Trois-Rivi\`eres. CP500 (QC)G9A 5H7, Canada}
\ead{grundlan@crm.umontreal.ca, post@crm.umontreal.ca}

\begin{abstract} Soliton surfaces associated with the $\CP$ sigma model are constructed using the Generalized Weierstrass  and the Fokas-Gel'fand formulas for immersion of 2D surfaces in Lie algebras. The considered surfaces are defined using continuous deformations of the zero-curvature representation of the model and its associated linear spectral problem.  The theoretical framework is discussed in detail and several new examples of such surfaces are presented. 
\end{abstract}

\section{Introduction}\label{intro} 
Integrable models and their continuous deformations under various types of dynamics have produced much interest and activity in several areas of mathematics, physics and biology. Furthermore soliton surfaces associated with integrable models, and  with the $\CP$ sigma model in particular, have been shown to play an essential role in many problems with physical applications (see e.g. \cite{Davbook, GPWbook, Landolffi2003, Raj2002, Saframbook}). The possibility of using a linear spectral problem (LSP) to represent a moving frame on the surface has yielded many new results concerning the intrinsic geometric properties of such surfaces (see e.g. \cite{BavMar2010, Bob1990, Bobbook, BobEit2000, Heleinbook2001, Kono1996, KonoTaim1996}). In this vein, it has  recently proved fruitful to extend such characterization of soliton surfaces via their immersion function in Lie algebras. The construction of such surfaces, related to the completely integrable $\CP$ sigma model,  has been accomplished by representing the equations of motion for the model as a conservation law which in turn provides a closed differential for the surface. This is the so called generalized Weierstrass formula for immersion (GWFI) \cite{ GoldGrund2009, GSZ2005, GrunYurd2009, KonLand1999}. 
Another method for the construction of such soliton surfaces makes use of the conformal invariance of the zero-curvature (or ZS-AKNS) representation of the model with respect to the spectral parameter. This immersion function is known as the Sym-Tafel formula for immersion \cite{Sym, Tafel} and a remarkable result, to be described later in this paper, is that the Sym-Tafel formula 
and the GWFI coincide in the case of finite action solutions of the $\CP$ sigma model. 

More generally, the LSP for the integrable model and its symmetries has been employed by Fokas and  Gel'fand \cite{FG} and later with Finkel and Liu \cite{FGFL}, to construct families of soliton surfaces. Recently, the authors have reformulated and extended the Fokas-Gel'fand immersion formula using the formalism of generalized vector fields and their action on jet space and have given the necessary and sufficient conditions for the existence and explicit integration of soliton surfaces in terms of the symmetry criterion for vector fields \cite{GrundPost2011a, GrundPost2011b, GrundPost2012}. 

The paper contains a survey of these recent results for the immersion of soliton surfaces,  particularly as applied to the completely integrable $\CP$ sigma model. The basics of the model are reviewed in section 2. Section 3 contains an exposition of the GWFI and its relation to the Sym-Tafel formula. The Fokas-Gel'fand formula for immersion, in the formalism of generalized vector fields,  and its associated surfaces are presented in section 4. Section 5 gives new examples of the application of such immersion formulas to the case of the $\mathbb{C}P^{1}$ sigma model.

\section{The $\CP$ sigma model: Definition and linear spectral problem}
Over the past decades, there has been significant progress in the study of general properties of the $\CP$ sigma model and the techniques for finding associated 2D-soliton surfaces immersed in multidimensional space. The most fruitful approach to this subject has been achieved through the descriptions of the model in terms of rank-one Hermitian projectors.   A matrix $P$ is said to be a rank-one Hermitian projector if 
\be P^2=P, \qquad tr(P)=1, \qquad P^\dagger=P.\ee
The target space of the projector $P$ is determined by a complex line in $\C$, i.e. by the one-dimensional vector function $f(\xi_+, \xi_-)$ given by 
\be\label{Pf} P=\frac{f\otimes f^\dagger}{f^\dagger\cdot f}, \ee
where $f$ is a mapping
\[ \mathbb{C} \supseteq\Omega \ni\xi_\pm =x\pm iy \mapsto f=(f_0, f_1, \ldots f_{N-1})\in \mathbb{C}^N \smallsetminus \{0\}.\]
In fact, the formula \eref{Pf} gives an isomorphism between the equivalence classes of $\CP$ and the set of rank-one, Hermitian projectors. The projector  formalism automatically encodes the scaling invariance of the vectors and ensures that the maps will be free from removable singularities which could occur in the unnormalized vector fields $f$ \cite{GoldGrund2010}. Furthermore, the equations of motion and other properties of the model take a compact form when written in the projector formalism. 

\subsection{The $\CP$ sigma model in the projector formalism}
In terms of a rank-one Hermitian projector $P$, the  Lagrangian density is given by
\be \label{lagrangian1} L(\Pb )=  tr(\partial_+ \Pb \bp \Pb ) \ee
and the action functional is 
\be \label{action} S(\Pb )=\int_{\mathbb{C}}{L}(\Pb )d\xi_+d\xi_-.\ee
Here the holomorphic and anti-holomorphic derivatives are defined as 
\[ \p=\frac{\partial}{\partial \xi_+}=\frac12\left(\frac{\partial }{\partial x}-i\frac{\partial }{\partial y}\right), \qquad \bp=\frac{\partial}{\partial \xi_-}=\frac12\left(\frac{\partial }{\partial x}+i\frac{\partial }{\partial y}\right).\]
The Euler-Lagrange (E-L) equation for the model is given by 
\be \label{EL} [\partial_+ \bp \Pb ,\Pb ]=\0,\ee
which is equivalent to a conservative form 
\be \label{ELconserv} \partial_+ [\bp \Pb ,\Pb ] +\bp[\partial_+ \Pb ,\Pb ]=\0.\ee
Here $\0$ is the zero matrix. 
Another important physical quantity of the model is the topological charge density
\be \label{chargedensity} q(\Pb )=tr(\Pb \p \Pb  \bp \Pb -\Pb \bp \Pb \p \Pb ),\ee
so called because its integral over the plane, the topological charge, is a total divergence and hence
\be\label{charge} Q(\Pb ) =\frac 1 \pi \int_{\mathbb{C}}q(\Pb ) d\xi_+ d\xi_-\ee
is an integer \cite{DinZak1980prop}. 

\subsection{Raising and lowering operators}
As was first proven by \cite{DinZak1980Gen, DinZak1980prop}, see also \cite{Zakbook}, any solution of the $\CP$ sigma model, defined on the extended complex plane, with finite action can be written as a raising operator acting on a holomorphic vector (or a lowering operator acting on an anti-holomorphic vector). In \cite{GrundPost2010}, the authors presented this proof in the projector formalism. In this context, the raising and lowering operators are given by \cite{GoldGrund2010}
\be\label{Pipm} \Pi_{\pm}(P)\equiv \Bigg{\lbrace} \begin{array}{cc} \frac{\partial_\pm P P\partial_\mp P}{tr(\partial_\pm P P\partial_\mp P)} & \partial_\pm P P\partial_\mp P \ne \0,\\
\0 &\partial_\pm P P\partial_\mp P = \0,\end{array} \ee
and have the property that a rank-one projector $P$ projects onto an equivalence class in $\CP$ with a holomorphic representative if and only if 
\be \label{PiXanal}\Pi_{-}P=\0.\ee
Analogously, $\Pi_{+}P=\0$ if and only if $P$ maps on to an equivalence class in $\CP$ with an anti-holomorphic representative. The operators $\Pi_\pm$ are called raising and lowering operators because they are contracting operators which map between solutions of the E-L equation \cite{GoldGrund2010}. They satisfy the following properties.  Suppose that a rank-one Hermitian projector $P$ is a solution of the E-L equation with finite action, then:

\begin{enumerate}
\item \label{E-Lrl} $\Pi_\pm P$ is either $\0$ or 
is itself a rank-one Hermitian projector solution of the E-L equation with finite action.

\item\label{rlinvesr} The raising and lowering operators are mutual inverses for solutions of the E-L equation, i.e. 
\be \Pi_\pm \Pi_\mp P=P, \qquad \mbox{ whenever }  \Pi_\mp P\ne \0.\ee
\item \label{rlcontract} The raising and lowering operators are mutually orthogonal
\be \Pi_+^jP\Pi_{+}^kP=\delta_{jk}\Pi_{+}^kP, \qquad j,k\in \mathbb{\mathbb{Z}},\ee
where the quantities $\Pi_{\pm}^{j}P$ are defined inductively by 
\begin{eqnarray*} &\Pi_{+}^0P=0, &\Pi_{+}^{-1}P=\Pi_-(P),\\
 &\Pi_{+}^{n+1}P=\Pi_+(\Pi_+^{n}P),  &\Pi_{+}^{n-1}P=\Pi_-(\Pi_+^{n}P). \end{eqnarray*}
\item The raising and lowering operators are contracting operators on $P$. That is, there exist natural numbers $j,k\in \mathbb{N}$ with $0\leq j+k\leq N-1$ such that 
\[ \Pi_{-}^{j+1}P=\0,\qquad \Pi_{+}^{k+1}P=\0.\]
\item The images of the finite set of rank-one projectors 
\be \label{Piset} \lbrace  \Pi_{-}^{j}P, \ldots \Pi_{-}P, P,\Pi_{+}P, \ldots \Pi_{+}^{k}P\rbrace\ee
form an orthogonal basis for the subspace $\mathbb{C}^{j+k+1}\subset\C.$ 
\end{enumerate}
If it is the case that $j+k+1<N$, then the $\CP$ model can be embedded in a $\mathbb{C}P^{j+k}$ model. Thus, for the remainder of the paper, it is assumed that $j+k+1=N$ and so
\[ \Pi_{-}^{j+1}P=\0,\qquad \Pi_{+}^{N-j}P=\0,\]
which in turn implies that $\Pi_{-}^{j}P$ is a holomorphic and $\Pi_{+}^{N-1-j}P$ and anti-holomorphic projector.

\subsection{The linear spectral problem}
 It is well known that the $\CP$ sigma model admits a linear spectral problem \cite{ZakMik1979, Zakbook}. Defining matrix functions
\be U^1\equiv\frac{2}{1+\lambda}[\p P, P], \qquad U^2\equiv\frac{2}{1-\lambda}[\bp P, P],\ee
the Euler-Lagrange (E-L) equations are  equivalent to
\be \label{EL} \Delta'\equiv [\p \bp P, P]=\bp U^1-\p U^2+[U^1,U^2]=0,\ee
which are exactly the compatibility conditions for the LSP of the form  
\be \label{cplsp} \p \Phi=U^1 \Phi, \qquad  \bp \Phi=U^2\Phi, \ee
where $\lambda$ is a complex spectral parameter. 

In is worth noting that, for the $\CP$ sigma model defined on Euclidean space, the wave function $\Phi$ can be explicitly integrated for an arbitrary solution of the E-L equation with finite action. The wave function $\Phi$ is given in terms of the set of rank-one Hermitian projectors \eref{Piset} by \cite{ GoldGrund2010, Zakbook}
\be\label{phi} \Phi=\left(\mathbb{I}+\frac{4\lambda}{(1-\lambda)^2}\sum_{j=1}^{k}\Pi_-^jP-\frac{2}{1-\lambda}P\right), \qquad \Pi_-^{k+1}P=\0,\ee
where $\mathbb{I}$ is the identity matrix on $\C.$  
If the spectral parameter $\lambda$ is purely imaginary, $\Phi$ is an element of the group $SU(N)$ and the inverse of the wave function is given by
\be\label{phiinv} \Phi^{-1}=\left(\mathbb{I}-\frac{4\lambda}{(1+\lambda)^2}\sum_{j=1}^{k}\Pi_-^jP-\frac{2}{1+\lambda}P\right).\ee  
The exact form of the wave function $\Phi$ will be important later in the construction of the soliton surfaces. 

\section{Generalized Weierstrass representation of surfaces associated with $\CP$ sigma models}
From the form of the E-L equation \eref{ELconserv}, it is possible to construct a closed, skew-Hermitian differential 
\be \label{dF} d\F=-i \left([\partial_+\Pb ,\Pb ]d \xi_+-[\partial_-\Pb ,\Pb ]d\xi_-\right).\ee 
Integrating the differential along a curve $\gamma \subset \mathbb{C}$ gives a surface immersed in the  $su(N)$ algebra associated with $\Pb ,$ a solution of the $\CP$ sigma model. These results are present in the following theorem \cite{GSZ2005}.
\begin{theorem} \label{Fthm} If $\Pb $ is a rank-1 Hermitian projector  solution of the Euler-Lagrange equations then there exists a surface $\F$ whose immersion is defined by 
\be \label{F} \F=-i\int_{\gamma} [\partial_+\Pb ,\Pb ]d \xi_+-[\partial_-\Pb ,\Pb ]d\xi_- , \qquad \F^\dagger =-\F\in su(N).\ee 
\end{theorem} 
The immersion function defined as in \eref{F} is called the Generalized Weierstrass formula for immersion (GWFI). Alternatively, the surface $F$ can be defined by its tangent vectors 
\be\label{dF} \frac{dF}{d\xi_+}=-i[\partial_+\Pb ,\Pb ]\qquad \frac{dF}{d\xi_-}=i[\partial_-\Pb ,\Pb ],\ee
whose compatibility conditions is exactly the E-L matrix equation \eref{ELconserv}.
The Killing form on  $su(N),$ 
\be \label{()} ({X},{ Y})=-\frac 12 tr({ X}\cdot {Y}), \qquad { X},{Y} \in su(N) \ee
gives an inner-product on the tangent vectors to a 2D-surface immersed in $su(N)$. 

With this metric, it is possible to show that the surfaces defined as in \eref{F} are conformally parameterized \cite{GrunYurd2009}.  The following theorem was proven in \cite{GrundPost2010}.
\begin{theorem}
 For a finite action solution, $P$, of the Euler-Lagrange equations \eref{EL}, the surface $\F$ defined by \eref{F} is conformally parameterized and the conformal factor in the first fundamental form is proportional to the Lagrangian density
\be I(F)=\frac12L(P)d\xi_+d\xi_-.\ee
 Thus, the area of the surface is given by the action functional of the model and,  in particular, the surface will have finite area.  
\end{theorem}

Other geometric quantities of the surface can be written in terms of the physical properties of the model. In complex coordinates, the Gaussian curvature becomes 
\be \label{KF} K(\F)=-\frac{2\partial_+\partial_-(ln(L(\Pb )))}{L(\Pb )}.\ee
The mean curvature vector, written in matrix form, is given by 
\be \label{mean} \H(\F)=\frac{-4i}{L(\Pb )}[\partial_+\Pb ,\partial_-\Pb ]\ee
and thus is traceless. 
The norm of $H$  can be written in terms of  the Lagrangian density and the topological charge density,
\bea\fl  (\H(\F),\H(\F))&=&\frac{8}{L(\Pb )^2}tr([\partial_+\Pb ,\partial_-\Pb ]^2)=\frac{4}{L(\Pb )^2}\left(L(\Pb )^2+ 3q(\Pb )^2\right).\eea
It is possible to express the Willmore functional in terms of the action, the Lagrangian density, and the topological charge density,
\be \label{W} W(\F)=\frac{1}2S(\Pb )+\frac{3}{2}\int_{\mathbb{R}^2}\frac{q(\Pb )^2}{L(\Pb )}d\xi_+d\xi_-.\ee
The Euler-Poincare character can be written terms of only the Lagrangian density
\bea \label{delta} \Delta(\F) &=& -\frac{1}{\pi} \int_{\mathbb{R}^2}\p \bp ln(L(\Pb ))d\xi_1d\xi_2.\eea

It was shown in \cite{GrunYurd2009} that the surfaces, defined as in theorem \ref{Fthm} can be integrated explicitly in terms of the set of orthogonal projectors \eref{Piset}.
\begin{theorem}\label{Fkthm}
Under the assumptions of theorem \ref{Fthm}, the surface  has an associated integer $k$ and holomorphic projector $\Pb _0$ so that the following holds
\be \label{Fk} \F=\F_k\equiv -i\left(\Pb _k+2\sum_{j=0}^{k-1}\Pb _j-\frac{1+2k}{N}\mathbb{I}\right), \qquad \Pb _\ell=\Pi^\ell_+\Pb _0.\ee 
\end{theorem}
The inverse formula for the projectors $\Pb_k$ in terms of the surface $F_k$ \eref{Fk} has been derived in \cite{GrunYurd2009}
\[ P_k=F_k^2-2i\left(\frac{1+2k}{N}-1\right)F_k-\frac{1+2k}{N}\left(\frac{1+2k}{N}-2\right)\mathbb{I}.\]
The exact integrated form of the surfaces \eref{Fk} allows one to obtain the minimal polynomials for the surfaces, dimensionality of the spanned subspaces of $\mathbb{R}^{N^2-1}\equiv su(N),$ and the angle between the immersion functions $F_k$ and $F_j$ of the 2D surfaces \cite{GrunYurd2009}.

\subsection{The Generalized Weierstrass  and the Sym-Tafel formulas for immersion}
An interesting fact is that the GWFI is equivalent to another formula for the immersion of 2D-surfaces in Lie algebras, namely the Sym-Tafel formula for immersion \cite{Sym, Tafel}. This immersion formula of a 2D surface is given by 
\be\label{Fst} F^{ST}=\alpha(\lambda)\Phi^{-1}\frac{\partial }{\partial \lambda} \Phi\in \g,\ee
whenever the tangent vectors 
\be \label{DFst} \p F^{ST}=a(\lambda)\Phi^{-1}\frac{\partial U^1 }{\partial \lambda} \Phi,\qquad \bp  F^{ST}=a(\lambda)\Phi^{-1}\frac{\partial U^2 }{\partial \lambda} \Phi\ee
are linearly independent. Here $a(\lambda) \in \mathbb{C}$ is an arbitrary function of $\lambda.$

From the specific form of $\Phi$ \eref{phi} and $\Phi^{-1}$ \eref{phiinv}, it is possible to directly compute the Sym-Tafel immersion formula as \cite{GrunYurd2009}
\be \label{Fstint} F^{ST}=\frac{2\alpha(\lambda)}{1-\lambda^2}\left(\Pb _k+2\sum_{j=0}^{k-1}\Pb _j-\frac{1+2k}{N}\mathbb{I}\right)\in su(N),\ee
which coincides with the GWFI, up to a multiplicative factor.

\section{Fokas-Gel'fand immersion formula for surfaces associated with $\CP$ sigma models}
In light of the results obtained for the Generalized Weierstrass formula for immersion and its equivalent form, the Sym-Tafel formula for immersion, it seems worthwhile to consider the larger class of immersion functions defined via the Fokas-Gel'fand formula for immersion. Such surfaces were recently considered by the authors in a series of papers \cite{GrundPost2011a, GrundPost2011b, GrundPost2012}. The necessary and sufficient conditions for the existence of such surfaces was formulated in terms of the symmetry criterion for generalized vector fields and their action on jet space \cite{Olver}. 

\subsection{The Fokas-Gel'fand immersion formula for surfaces}
The considered surfaces are defined via infinitesimal deformations of the zero-curvature condition 
\be \label{Delta} \Delta[u]= u^1_2-u^2_1+\left[u^1, u^2\right]=0,\ee
with independent variables $\xi_{i}, \ i=1,2$ and dependent matrix variables $u^\alpha,$ $\alpha=1,2$ which take their values in a Lie algebra $\g.$ In the case of the $\CP$ sigma model defined on Euclidean space, the independent variable  take the form $\xi+=\xi_1$ and $\xi_-=\xi_2$ and the Lie algebra is $\g=su(N).$  The PDE \eref{Delta} can be considered  as a function on the jet space  $M\equiv(\xi_1, \xi_2, u^\alpha, u^\alpha_J),$  where  the derivatives of $u^\alpha$ are given by 
 \be \frac{\partial ^n}{\partial \xi_{j_1}\ldots \partial \xi_{j_n}} u^\alpha\equiv u^\alpha_J, \qquad J=(j_1, \ldots j_n), \quad j_i=1,2, \ |J|=n.\ee 
 Define the set of smooth functions on the jet space $M$ to be  $\mathcal{A}\equiv C^\infty (M).$ The abbreviated notation $f(\xi_1, \xi_2, u^\alpha, u^\alpha_J)\equiv f[u]\in\mathcal{A}$ will be used. The zero-curvature condition $\Delta[u]=0$ can be realized as the compatibility conditions for a wave function $\Psi[u]$ defined by its tangent vectors
\be\label{lsppsi} D_\alpha\Psi=u^\alpha\Psi, \qquad \alpha=1,2,\ee
 where the total derivatives $D_i$ are defined as 
 \be D_i=\frac{\partial}{\partial \xi_i}+u^{\alpha, j}_{J,i}\frac{\partial}{\partial u^{\alpha, j}_J}, \qquad \alpha,i=1,2.\ee
 Here $u^{\alpha, j}$ are the components of  $u^\alpha$ in a basis for $\g,$ i.e.  $u^{\alpha}=u^{\alpha, j}e_j$ with $j=1...n.$ 
 
Suppose now that it is possible to parameterize the matrices  $u^\alpha \rightarrow U^\alpha([\theta], \lambda)\in \g$ in terms of some set of dependent variables $\theta^j(\xi_+, \xi_-),$ which depend on  the independent variables  $\xi_\pm.$ The potential matrices $U^\alpha([\theta], \lambda)$ are assumed to depend on a spectral parameter $\lambda$ in such a way  that the zero-curvature condition is equivalent to an integrable PDE in dependent variables $\theta^j$ which is independent of the spectral parameter
\be\label{Delta'}  \Delta'[\theta]=D_1U^2-D_2U^1+\left[U^1, U^2\right]=0 .\ee 
In the case of the $\CP$ sigma model, the scalar functions $\theta^j$ represent the components of the projector $P$.  

Define a new jet space  $N$ to be the jet space associated with $\theta^j$ and its derivatives and $\mathcal{B}$ to be the space of smooth functions on this jet space, also possibly depending  on the spectral parameter $\lambda.$ The elements of $\mathcal{B}$ are denoted $f(\lambda, \xi_+, \xi_-, \theta^j, \theta^j_J)=f([\theta], \lambda) \in \mathcal{B}.$ Let $\tau $ be the mapping between the function spaces   $\mathcal{A} $ and $\mathcal{B}$ defined by taking   $\tau(u^\alpha_J)= D_J\left(U^\alpha([\theta], \lambda)\right):$ 
 \be \begin{array}{ccc} &\tau &\\
 u^\alpha\in \g &\rightarrow& U^\alpha([\theta])\in \g \\
 \downarrow & & \downarrow\\
 & \tau &\\
 \mathcal{A}=C^\infty(M) & \rightarrow &\mathcal{B}=C^\infty(N\times \mathbb{C}). \end{array} \nonumber\ee
 Because of the structure of jet space, the mapping $\tau$ commutes with total derivatives. 
 Under the mapping $\tau$, the LSP becomes
 \be \label{LSPphi} D_\alpha\Phi=U^\alpha\Phi, \qquad \alpha=1,2,\ee
 where the wave function is given by $\Phi([\theta], \lambda)=\tau(\Psi[u])$ and potential matrices $U^\alpha([\theta],\lambda)=\tau (u^\alpha).$
 
 This  spectral problem was used by Fokas and Gel'fand in \cite{FG} to construct immersion functions for surfaces in Lie algebras whose Gauss-Mainardi-Codazzi (GMC) equations are given by 
  \be\label{AB'} D_1 A-D_2  B+[A,U^2]+[U^1,B]=0,\ee
an infinitesimal deformation of \eref{Delta'}. Here $A$ and $B$ are elements of the Lie algebra $\g$ with components in the function space $\mathcal{B}.$ The immersion functions are defined by their tangent vectors
\be\label{DF} D_1F=\Phi^{-1} A\Phi, \qquad D_2F=\Phi^{-1} B\Phi.\ee
It is straightforward to verify that the compatibility conditions of \eref{DF} are satisfied whenever the LSP \eref{LSPphi} and the GMC equations \eref{AB'} hold. 
 
The possible forms of such matrices $A$ and $B$ have been studied by several authors, e.g. \cite{Cies1997, FGFL, GrundPost2011a, GrundPost2011b, GrundPost2012}.  In this case, there are three types of admissible symmetries:
\begin{itemize}
\item   generalized symmetries of the zero-curvature condition itself $\Delta[u]=0$,
\item  generalized symmetries of the corresponding integrable PDE $\Delta'[\theta]=0$,
\item  invariance of the integrable PDE with respect to a conformal transformation in the spectral parameter. This is the Sym-Tafel formula for immersion. 
\end{itemize}
 
 In order to construct such symmetries, it is expedient to consider generalized fields on jet space. A vector field $\vec{v}_Q$  defined on the jet space $M$ in evolutionary form is given by 
\be\label{vQ} \vec{v}_Q=Q^{\alpha,j}[u] \frac{\partial}{\partial u^{\alpha,j}}.\ee
Note that $Q^\alpha\equiv Q^{\alpha, j}e_j$ is an element of the Lie algebra $\g$. The prolongation of $\vec{v}_Q$ is defined in the standard way
 \be\label{prv} pr\vec{v}_Q=\vec{v}_Q+D_J(Q^{\alpha,j}[u]) \frac{\partial}{\partial u^{\alpha,j}_J}.\ee
  A vector field $\vec{w}_\Q$ defined on the jet space $N$ in evolutionary form is denoted by 
\be \vec{w}_\Q=\Q^{j}[\theta] \frac{\partial}{\partial \theta^j},\qquad \Q^j[\theta]\in \mathcal{B} .\ee
The notation $\Q$ is used to distinguish the characteristics of a vector field $\vec{v}_Q$ on $M$ from a vector field $\vec{w}_\Q$ on $N$. The prolongation of $\vec{w}_\Q$ is defined in the standard way by analogy with \eref{prv}.

A generalized vector field $\vec{v}_Q$ on jet space $M$ is said to be a generalized symmetry of the nondegenerate PDE  $\Delta[u]=0$ if and only if \cite{Olver}
\be\label{prD} pr\vec{v}_Q\left(\Delta[u]\right)=0,\qquad \mbox{ whenever }\Delta[u]=0.\ee
From the definition of the prolongation \eref{prv}, it is immediate to verify that \eref{prD} is equivalent to 
\be \label{Q1Q2} D_2Q^1-D_1Q^2+[Q^1,u^2]+[u^1, Q^2]=0\ee
Thus, the $\g$-valued functions on jet space $N$
\be\label{ABv} A=\tau(Q^1), \qquad B=\tau(Q^2),\ee
satisfy \eref{AB'} whenever $\vec{v}_Q$ is a symmetry of $\Delta[u]=0$. 

Similarly, a generalized vector field $\vec{w}_\Q$ on jet space $N$ is said to be a generalized symmetry of the nondegenerate PDE $\Delta'[\theta]=0$ if and only if 
\be\label{prwD} pr\vec{w}_\Q\left(\Delta'[\theta]\right)=0,\qquad \mbox{ whenever }\Delta'[\theta]=0.\ee
From the definition of the prolongation \eref{prv}, it is immediate to verify that \eref{prD} is equivalent to 
\be \label{prwDexp} D_2(pr\vec{w}_\Q U^1)-D_1(pr\vec{w}_\Q U^2)+[pr\vec{w}_\Q U^1,U^2]+[U^1, pr\vec{w}_\Q U^2]=0.\ee
Thus, the $\g$-valued functions on jet space $N$
\be \label{ABw} A=pr\vec{w}_\Q U^1, \qquad B=pr\vec{w}_\Q U^2,\ee
satisfy \eref{AB'} whenever $\vec{w}_\Q$ is a symmetry of $\Delta'[\theta]=0$.

The third possible choice of $A$ and $B$ is associated with the invariance of $\Delta'[\theta]=0$ with respect to conformal symmetries of the spectral parameter $\lambda.$ In this case, the choice of matrices $A$ and $B$ are given by 
\[A=a(\lambda)\frac{\partial}{\partial \lambda} U^1, \qquad  B=a(\lambda)\frac{\partial}{\partial \lambda} U^2,\]
which satisfy \eref{AB'} whenever $\Delta'[\theta]=0$ is independent of the spectral parameter. With this choice of $A$ and $B$, the tangent vectors given by \eref{DF} coincide with those for the Sym-Tafel formula for immersion \eref{DFst}. Thus, the surfaces take the integrated form given in \eref{Fst}.

\subsection{Integrated form of the surfaces immersed in Lie algebras}
As in the case of the Sym-Tafel formula for immersion, it is often possible to give explicit integrated forms for the immersion function for the surfaces. 
As proven in \cite{GrundPost2012}, any symmetry of the zero-curvature condition $\Delta[u]=0$ can be written in terms of a gauge function $S[u]$ and the associated immersion function is given by 
\be \label{Fs}F=\Phi^{-1}\tau\left( S[u]\right)\Phi\in\g.\ee
For example, the zero-curvature condition $\Delta[u]=0$ is invariant under a scaling of the independent and dependent variables associated with the vector field $\vec{v}_Q$ with
\be\label{symm1} Q^1= D_1\left(\xi_1u^1\right)+\xi_2D_2\left(u^1\right), \qquad Q^2=\xi_1D_1\left(u^2\right)+D_2\left(\xi_2u^2\right).\ee
The gauge associated with this symmetry is \cite{GrundPost2012}
 \[S[u]=\xi_1u^1 +\xi_2u^2\]  and the corresponding surface is
\be\label{FvQ} F=\Phi^{-1}\left( \xi_1U^1+\xi_2U^2\right)\Phi\in \g.\ee
It is straightforward to verify that the tangent vectors to this surface  are given as in \eref{DF} with 
\[ \fl A=\tau(Q^1)=D_1\left(\xi_1U^1\right)+\xi_2D_2\left(U^1\right),\qquad B=\tau(Q^2)=\xi_1D_1\left(U^2\right)+D_2\left(\xi_2U^2\right).\]

As a second example, consider the following generalized symmetry of $\Delta[u]=0$ whose generalized vector field $\vec{v}_Q$ has characteristics
\be\label{symm2} \fl Q^1=u^1_{11}+u^1_{22}+[u^1_1,u^1]+[u^1_2,u^2], \qquad 
Q^2=u^2_{11}+u^2_{22}+[u^2_1,u^1]+[u^2_2,u^2].\ee
The gauge associated to this symmetry is given by \[S[u]=u^1_1+u^2_2\] and the corresponding surface takes the form
\be \label{FvQg} F=\Phi^{-1} \left( D_1 U^1+D_2U^2\right)\Phi\in \g.\ee
As in the previous case, the tangent vectors to this surface are given as in \eref{DF} with 
\begin{eqnarray*} A&=&\tau(Q^1)=D_1^2U^1+D_2^2U^1+[D_1U^1,U^1]+[D_2U^1,U^2] ,\\
 B&=&\tau(Q^2)=D_1^2U^2+D_2^2U^2+[D_2U^2,U^2]+[D_1U^2,U^1] .\end{eqnarray*}

There also exist soliton surfaces associated with generalized symmetries of the integrable model, such as in the case of conformal symmetries of the $\CP$ sigma model.  As was proven in \cite{GrundPost2011a},  the surface associated with such symmetries, i.e. with $A$ and $B$ of the form \eref{ABw}, can be integrated explicitly as 
\be \label{Fw} F=\Phi^{-1} pr\vec{w}_\Q\Phi\in \g,\ee
whenever $\vec{w}_\Q$ is a generalized symmetry of the LSP in the sense that 
\be \label{symlsp} pr\vec{w}_Q\left(D_\alpha \Phi-U^\alpha\Phi\right)=0 \mbox{ whenever } D_\alpha \Phi-U^\alpha\Phi=0.\ee
In general, the requirement \eref{symlsp} is difficult to verify since it requires a knowledge of the integrated form of the wave function $\Phi.$ However, in the case of finite action solutions of  the $\CP$ sigma model, the wave function is known explicitly \eref{phi} and it has been directly shown that conformal symmetries of the independent variables are also symmetries of the LSP \cite{GrundPost2011a}. Thus, the surface $F$ as in \eref{Fw} takes the particular form 
\be \label{FwC} F=\Phi^{-1}\left( g(\xi_+)U^1+\overline{g}(\xi_-)U^2\right)\Phi\in su(N),\ee
associated with vector field 
\[ \vec{w}=-g(\xi_+)\p -\overline{g}(\xi_-)\bp,\]
which is a conformal symmetry of the the $\CP$ sigma model. 

\section{Soliton surfaces associated with the $\mathbb{C}P^{1}$ sigma model}
To illustrate these theoretical considerations, examples of soliton surfaces associated with the $\mathbb{C}P^{1}$ sigma model are presented. In this model, the only solutions with finite action are holomorphic and anti-holomorphic projectors. Nevertheless, as will be demonstrated below, there are various forms of surfaces which can be constructed in this manner. 

For these simple examples, the  holomorphic projector is chosen as the following matrix, based on the Veronese sequence $f=(1, \xi_+)$,
\be \Pb _0=\frac{1}{(1+|\xi|^2)^2} \left[\begin{array} {cc} 
1 & \xi_- \\
\xi_+ &|\xi|^2 \end{array}\right], \qquad |\xi|^2\equiv \xi_+\xi_-. \ee
The wave function in the LSP for this model is simply
\be \Phi_0=\left(\mathbb{I}-\frac{2}{1-\lambda}P\right).\ee
The surface corresponding to the Sym-Tafel formula for immersion is written as 
\[ F^{ST}=\frac{i}{(1+|\xi|^2)} \left[ \begin {array}{cc} 1&\xi_-\\\noalign{\medskip}\xi_+&\xi_+\xi_-\end {array}
 \right] ,
\]
where the scaling factor is chosen as $i(1-\lambda^2)/2$ so as to coincide with the GFWI. This surface has constant positive Gaussian and mean curvature given by
\[ K=H=4,\]
and so is in fact a sphere, as demonstrated in figure 1. The surfaces are represented in terms of their components in the following basis for $su(2)$
\be \label{basis} e_1=\left[\ba{cc}0&i\\ i&0 \ea \right], \quad e_2=\left[\ba{cc}0&-1\\ 1&0 \ea \right], \quad e_3=\left[\ba{cc}i&0\\ 0&-i \ea \right].\ee

The surface associated with the (scaling) point symmetry of the zero-curvature condition $\vec{v}_Q $ \eref{vQ} has the integrated form 
\begin{eqnarray*} \fl F^{P}&=&\Phi^{-1}\left(\xi_1U^1+\xi_2U^2\right)\Phi\\
\fl &=&
\frac{1}{(1-\lambda^2)(1+|\xi|^2)^2} \left[ \begin {array}{cc} 4\,\lambda\,\xi_+\xi_-&2\, \left((\lambda+1)
\,\xi_+\xi_-+1 -\lambda\right) \xi_-\\\noalign{\medskip}2\, \left( (\lambda-1)\,\xi_+\xi_-
-1-\lambda \right) \xi_+&-4\,\lambda\,\xi_+\xi_-\end {array} \right]. \end{eqnarray*}
This surface, $F^{P}$, also has constant positive Gaussian and mean curvatures
\[ K=-4\lambda^2, \qquad H=-4i\lambda,\qquad i\lambda\in \mathbb{R}\]
though the surface is not a sphere since it has a boundary. A graph of the surface is shown in figure 1. 
\begin{figure}\label{fig1}
\begin{center}$
\begin{array}{cc}

\includegraphics[width=2.5in]{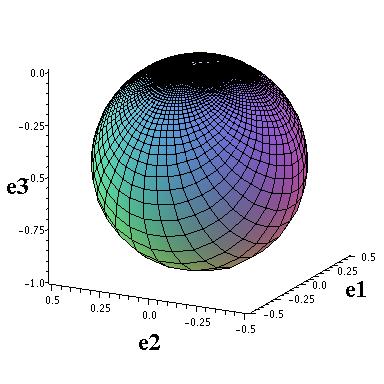} & \includegraphics[width=2.5in]{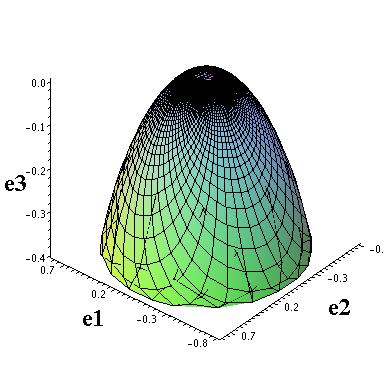}\\
F^{ST} &F^{P}\end{array}$
\caption{Surfaces $F^{ST}$ and $F^{P}$ with $\lambda=i/2$ and $\xi_\pm=x\pm iy$ with $x$ and $y \in [-20,20]$. The axes indicate the components of the immersion function in the basis \eref{basis}  }
\end{center}
\end{figure}

In the case of the surface associated with the generalized symmetry of the zero-curvature condition \eref{FvQg}, the immersion function is given by
\begin{eqnarray*}\fl  F^{G}&=&\Phi^{-1}\left(D_1U^1+D_2U^2\right)\Phi\\
\fl &=&
\frac{4}{(1-\lambda^2)(1+|\xi|^2)^3} \left[ \begin {array}{cc} -\lambda(\,{\xi_+}^{2}+\,{\xi_-}^{2})+{\xi_+}^{2}-
{\xi_-}^{2}&-(\lambda+1){\xi_-}^{3}+(\lambda-1)\xi_+\\\noalign{\medskip}
(1-\lambda)\,{\xi_+}^{3} +(1+\lambda)\xi_-&+\lambda ({\xi_+}^{2}+{\xi_-}^{2})-{\xi_+}^{2}+
{\xi_-}^{2}\end {array} \right]. \end{eqnarray*}
The surface associated with conformal symmetry of the E-L equation \eref{FwC} with $g(\xi_+)=1$ is given by 
 \begin{eqnarray*}\fl  F^{C}&=&\Phi^{-1} \left(U^1+U^2\right)\Phi\\
 \fl &=& \frac{1}{(1-\lambda^2)(1+|\xi|^2)^2}\left[\ba{cc} \xi_--\xi_++\lambda(\xi_+-\xi_-) & (1+\lambda)\xi_-^2+1-\lambda \\
(\lambda-1)\xi_-^2-1-\lambda &  \xi_+-\xi_--\lambda(\xi_+-\xi_-) \ea \right]
.\end{eqnarray*}
In both cases the Gaussian and mean curvatures are rational functions though they are too involved to  write out in an illustrative fashion. Their graphs are given in figure 2. 
\begin{figure}\label{fig1}
\begin{center}$
\begin{array}{cc}

\includegraphics[width=2.5in]{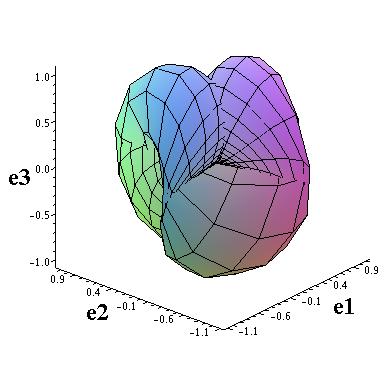} & \includegraphics[width=2.5in]{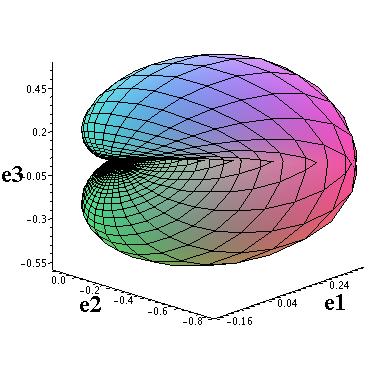}\\
F^{G} &F^{C}\end{array}$
\caption{Surfaces $F^{G}$ and $F^{C}$ with $\lambda=i/2$ and $\xi_\pm=x\pm iy$ with $x$ and $y \in [-20,20]$. The axes indicate the components of the immersion function in the basis \eref{basis}  }
\end{center}
\end{figure}

\section{Future outlook}
The approach for constructing soliton surfaces immersed in Lie algebras described in this paper has proven to be very effective in revealing geometric properties of investigated classes of surfaces. The future objective is to extend it to more general sigma models and study surfaces related to the complex Grassmannian models and possibly to models associated with octonion geometry. Since the field theoretic formulation of soliton surfaces by the Euler-Lagrange equation naturally involves the variation principle, it also leads in a straightforward way to geometric objects like geodesics, harmonic maps, minimal surfaces and the like. It also relates closely to the formalism of completely integrable systems, e.g. Lax pairs, conservations laws and Hamiltonian structures. So, it is apparent that the methods bring a strong unifying potential to the prospective research.

\ack S Post would like to thank the organizers, Dieter Schuch and Michael Ramek, for their kind invitation and support in attending the conference.   The research reported in this paper is supported by NSERC of Canada. S Post acknowledges a postdoctoral fellowship provided by the Laboratory of Mathematical Physics of the Centre de Recherches Math\'ematiques, Universit\'e de Montr\'eal.

\end{document}